\documentclass[]{spie}  

\usepackage{amsmath,amsfonts,amssymb}
\usepackage{graphicx}
\usepackage{ulem}

\title{Optical Design of PICO, a Concept for a Space Mission to Probe Inflation and Cosmic Origins}

\author[a\dag]{Karl Young}      
\author[b]{Marcelo Alvarez}  
\author[c]{Nicholas Battaglia}  
\author[d]{Jamie Bock}       
\author[e]{Julian Borrill}  
\author[f]{David Chuss}  
\author[g]{Brendan Crill}    
\author[h]{Jacques Delabrouille}  
\author[i]{Mark Devlin}  
\author[j]{Laura Fissel}  
\author[k]{Raphael Flauger} 
\author[l]{Daniel Green}  
\author[g]{Kris Gorksi}  
\author[a]{Shaul Hanany} 
\author[m]{Richard Hills} 
\author[n]{Johannes Hubmayr} 
\author[o]{Bradley Johnson}  
\author[c]{Bill Jones}  
\author[p]{Lloyd Knox}  
\author[q]{Al Kogut}  
\author[g]{Charles Lawrence}  
\author[r]{Tomotake Matsumura} 
\author[g]{Jim McGuire}  
\author[s]{Jeff McMahon}  
\author[g]{Roger O'Brient} 
\author[a]{Clem Pryke}  
\author[g]{Brian M. Sutin}  
\author[a]{Xin Zhi Tan}  
\author[g]{Amy Trangsrud}  
\author[a]{Qi Wen}  
\author[t]{Gianfranco de Zotti}  

\affil[a]{University of Minnesota, USA}
\affil[b]{University of California Berkeley, USA}
\affil[c]{Center for Computational Astrophysics, Flatiron Institute, USA}
\affil[d]{California Institute of Technology, USA}
\affil[e]{Lawrence Berkeley National Laboratory, USA}
\affil[f]{Villanova  University, USA}
\affil[g]{Jet Propulsion Laboratory, California Institute of Technology, USA}
\affil[h]{Laboratoire AstroParticule et Cosmologie and CEA/DAP, France}
\affil[i]{University of Pennsylvania, USA}
\affil[j]{National Radio Astronomy Observatory, USA}
\affil[k]{University of California San Diego, USA}
\affil[l]{University of Toronto, Canada}
\affil[m]{Cavendish Laboratory, University of Cambridge, UK}
\affil[n]{National Institute of Standards and Technology, USA}
\affil[o]{Columbia University, USA}
\affil[p]{University of California Davis, USA}
\affil[q]{Goddard Space Flight Center, USA}
\affil[r]{Kalvi IPMU, University of Tokyo, Japan}
\affil[s]{University of Michigan, USA}
\affil[t]{Osservatorio Astronomico di Padova, Italy}

\authorinfo{$^\dag$kyoung@astro.umn.edu}

\begin{document} 
\maketitle

\begin{abstract}

The Probe of Inflation and Cosmic Origins (PICO) is a probe-class mission concept currently under study by NASA.  
PICO will probe the physics of the Big Bang and the energy scale of inflation, constrain the sum of neutrino masses, 
measure the growth of structures in the universe, and constrain its reionization history by making full sky maps of the 
cosmic microwave background with sensitivity 80 times higher than the {\it Planck} space mission. With bands at 
21-799~GHz and arcmin resolution at the highest frequencies, PICO will make polarization maps of Galactic synchrotron 
and dust emission to observe the role of magnetic fields in Milky Way's evolution and star formation. 
We discuss PICO's optical system, focal plane, and give current best case noise estimates. The optical 
design is a two-reflector optimized open-Dragone design with a cold aperture stop. It gives a diffraction limited
field of view (DLFOV) with throughput of 910 cm$^{2}$sr at 21 GHz. The large 82 square degree DLFOV 
hosts 12,996 transition edge sensor bolometers distributed in 21 frequency bands and maintained at 0.1~K. 
We use focal plane technologies that are currently implemented on operating CMB instruments including three-color 
multi-chroic pixels and multiplexed readouts. To our knowledge, this is the first use of an open-Dragone design 
for mm-wave astrophysical observations, and the only monolithic CMB instrument to have such a broad frequency 
coverage. With current best case estimate polarization depth of 0.65~$\mu$K$_{\rm CMB}$-arcmin over the entire 
sky, PICO is the most sensitive CMB instrument designed to date. 

\end{abstract}


\section{INTRODUCTION}
\label{sec:intro}

Over the last decade NASA's astrophysics division has funded design and construction of space missions that are either Explorer-class, 
with cost cap of up to \$250M or Flagship-class that cost above \$1B. 
To study the science opportunities available at intermediate costs, NASA initiated studies of Probe-class missions with cost window 
between \$400M and \$1B.  We are conducting one these studies for a mission called the Probe of Inflation and Cosmic Origins (PICO). 
A paper by Sutin et al.\cite{brian_spie}\ in these proceedings gives an overall review of PICO and the scientific motivation.  
This paper describes the design of the telescope and focal plane and gives 
our current best estimates for the sensitivity of the instrument. The mission study is not complete; the final report 
is due in December 2018. Therefore, the quantitative assessments we provide for component temperatures and detector
noise levels are temporary in nature and subject to revision. 
Even so, the design is fairly mature and we do not expect significant changes. 
Values in this paper are current best estimates and don't represent finalized mission requirements. 


\section{MISSION AND SPACECRAFT}
\label{sec:spacecraft}

PICO will conduct scientific observations for five years from an orbit around the Earth-Sun L2 Lagrange point. The design had
21 bands centered at 21--799~GHz.  The spacecraft design impacted 
the optical design and sensitivity in two primary ways; volume constraints limited the physical size of the telescope and optical component 
temperatures impacted noise levels.  

\begin{figure} [ht]
\begin{center}
\includegraphics[height=9cm]{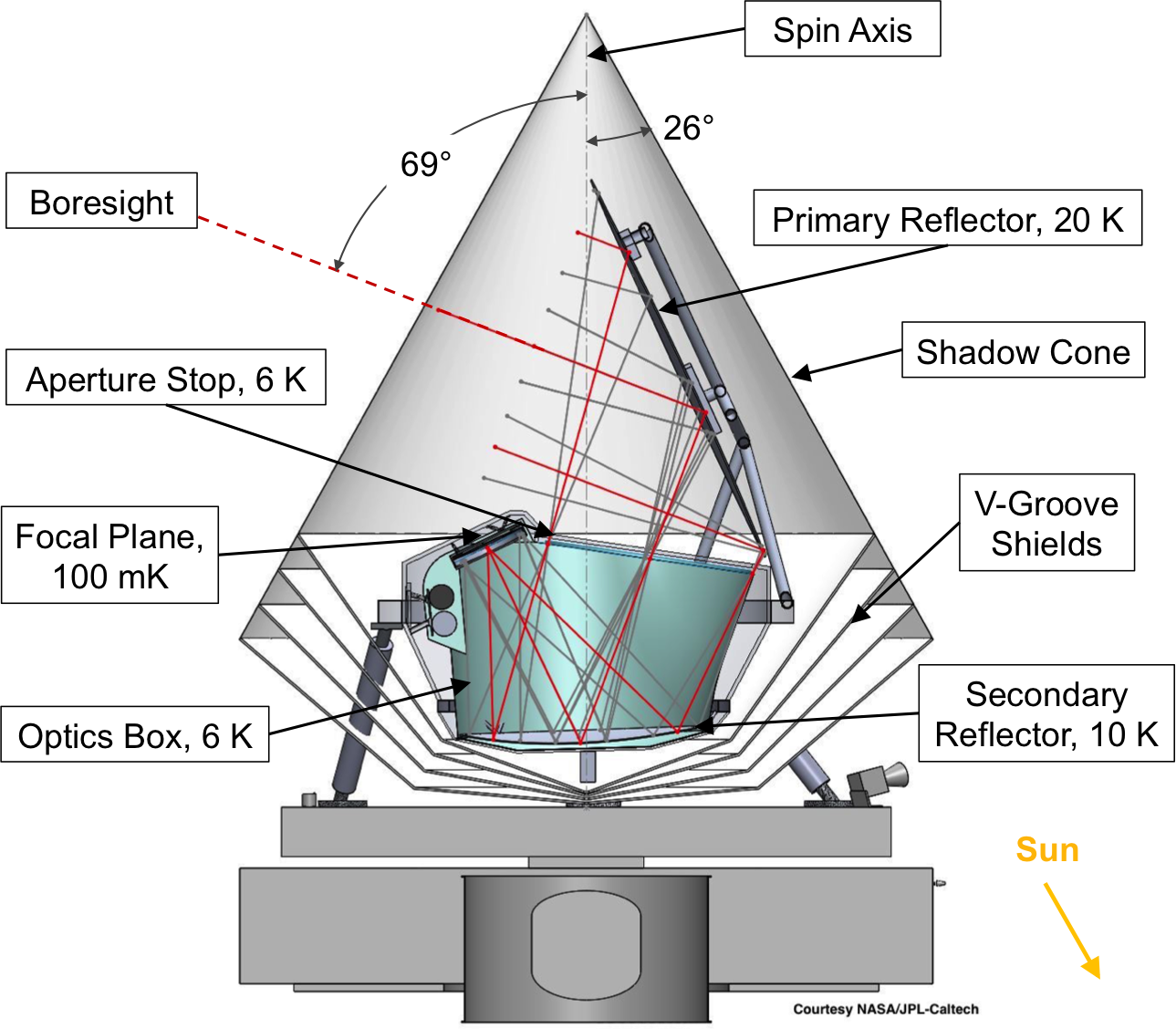}
\end{center}
\caption { \label{fig:cad} 
The spin axis of the satellite precesses around the satellite-sun axis (orange arrow) with an angle of 26 deg.
This precession defines the shadow cone shown in light gray.
}
\end{figure}

The maximum diameter of the spacecraft is limited by the 4.6~m largest diameter that a SpaceX's Falcon 9 launch vehicle can 
carry.  This diameter limits the V-groove shields' size, which, along with the scan strategy, defines the `shadow cone' in Figure~\ref{fig:cad}.  
The shadow cone is the volume protected from solar illumination, and all optical components are contained within it. The shadow cone and 
inner V-grooves defines an available volume for the telescope.  We opted not to use deployable shields as they present added cost 
and risk which outweighs the benefits. The thermal model indicates the temperatures of the optical elements as given in Figure~\ref{fig:cad}. 
The primary reflector is passively cooled, while the optics box, secondary reflector, and focal plane are actively cooled; Sutin~et~al.\cite{brian_spie}
gives more details of the thermal system. 

\section{OPTICAL SYSTEM}
\label{sec:optics}

The choice of telescope design is driven by a combination of science
requirements and the physical limits discussed in Section~\ref{sec:spacecraft}.  The science requirements are: a large diffraction 
limited field of view (DLFOV)\footnote{ We consider an area in the FOV diffraction limited when the Strehl ratio is 
larger than 0.8.} sufficient to support $\mathcal{O}(10^4)$ detectors, arcminute resolution at 800~GHz, low 
instrumental polarization, and low sidelobe response. Additionally, 
the transition edge sensor bolometers baselined for PICO require a telecentric focal plane which is sufficiently flat that it 
can be tiled by 10~cm detector wafers without reduction in optical quality. 

\begin{figure} [ht]
\begin{center}
\includegraphics[height=7.5cm]{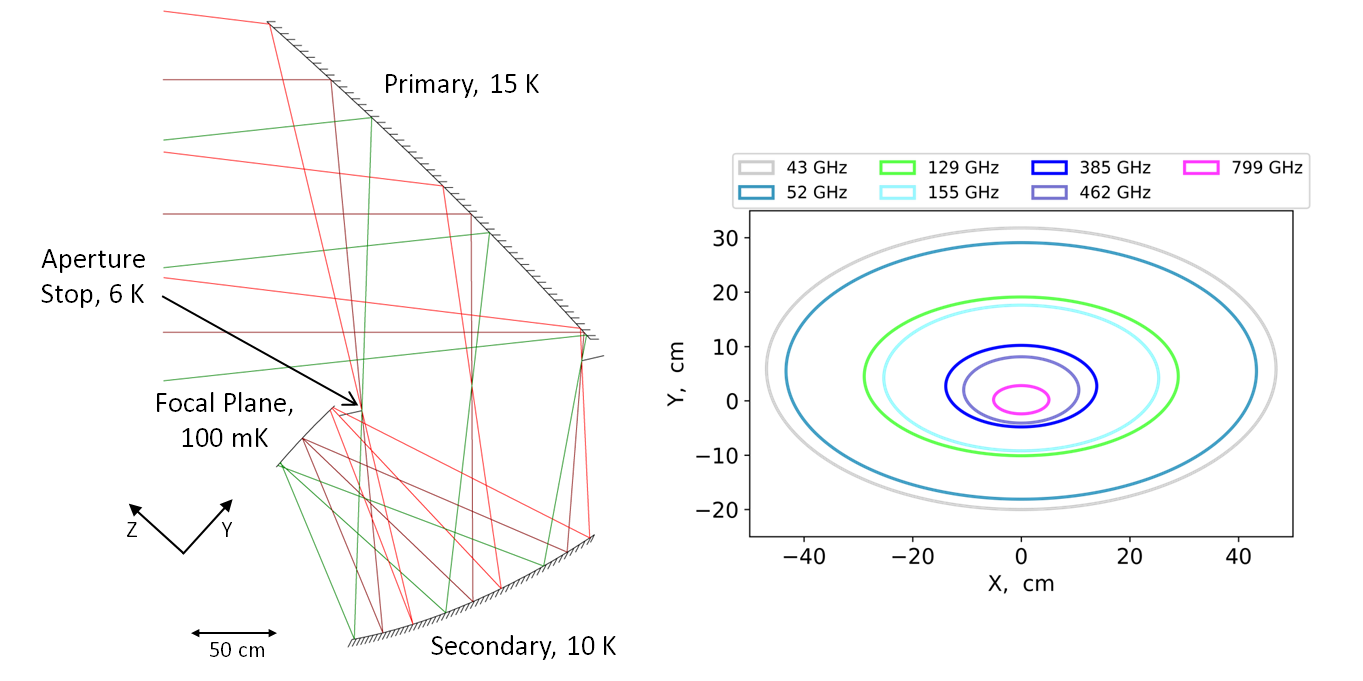}
\end{center}
\caption { \label{fig:ray} \label{fig:strehl} 
Raytrace (left) and Strehl~$=0.8$ contours (right) for the PICO optical design. }
\vspace{0.2in}
\end{figure}

To increase aperture efficiency and reduce sidelobes we concentrated our investigation around off-axis optical designs.
More than 30 years ago Dragone analyzed the 
performance of several off-axis systems and found solutions with low cross-polarization at the center of the field 
of view and with astigmatism, or astigmatism and coma, canceled to first order.\cite{dragone,dragone_coma,dragone1983} These
systems also have no cross-polarization at the center of the field of view. 
A number of recent CMB instruments used off-axis systems, and several 
began the design optimization with systems based on designs by
Dragone\cite{planck2000_optics,ACT2011_optics,SPT2008_optics,core2018_inst,LB2016_optics,parshley_ccat_spie}. 
For PICO we began 
the optimization with a two-reflector Dragone system that, to our knowledge, has not been implemented in CMB instruments before. 
We call it an `open-Dragone' because of its overall geometry and in contrast to the widely used `cross-Dragone', see Figure~\ref{fig:ray}. 
We used a 1.4~m entrance aperture as this aperture diameter satisfies the science requirements.

\begin{figure} [ht]
\begin{center}
\includegraphics[height=6.5cm]{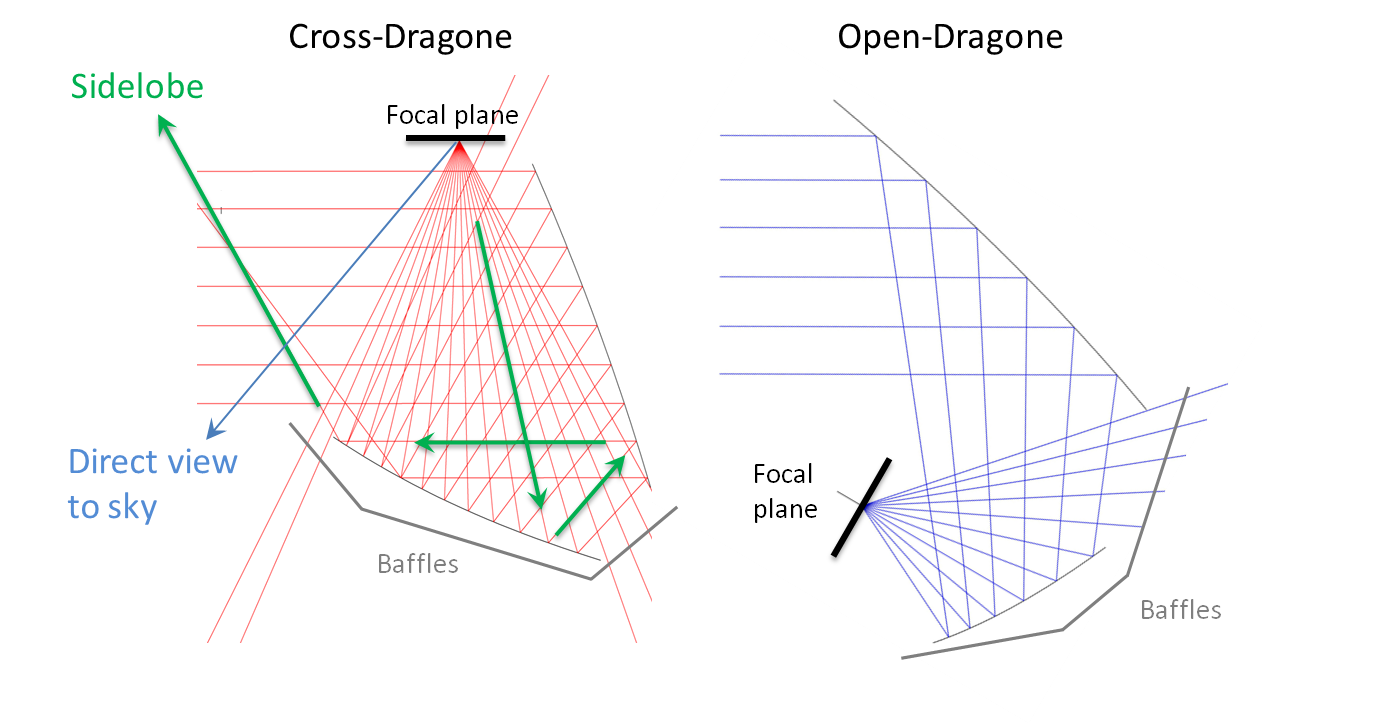}
\end{center}
\caption { \label{fig:sidelobes} 
Comparison of sidelobes for a cross-Dragone (left) and an open-Dragone (right).  
Rays were traced from the center of the focal plane toward the sky.
For both systems spillover around the secondary was straightforward to mitigate with absorptive baffles.  
However, the rays labeled `sidelobe' and `direct view to 
sky' in the cross-Dragone system presented added challenges. Those challenges could be mitigated---with a long forebaffle or larger $F$-number (see
for example Matsumrua et al.\cite{LB2016_optics})---but
doing so increased the overall physical size of the system, which was problematic in the PICO case.}
\end{figure}

We considered two additional Dragone systems, a Gregorian Dragone and a cross-Dragone, and compared the 
relative performance of all three systems in terms of DLFOV, compactness, and rejection of sidelobes.  
Compared to the open-Dragone, the Gregorian had half the DLFOV for the same $F$-number and could not  
support $\mathcal{O}(10^4)$ detectors. It was therefore rejected.  
The cross-Dragone had roughly $4\times$ the DLFOV of the open-Dragone, 
but was more difficult to pack inside the spacecraft volume while avoiding the known sidelobes shown in Figure~\ref{fig:sidelobes}.
We found that the largest cross-Dragone which met the PICO volume constraints had 
a 1.2~m aperture and an $F$-number of 2.5, while the largest open-Dragone aperture was 1.4~m with an $F$-number of 1.42. The larger 
$F$-number of the cross-Dragone system implied a larger physical focal plane, and therefore higher mass and cost, 
for the same number of pixels. 
For the PICO case, we concluded that the advantages of a low $F$-number and easily baffled sidelobes made the open-Dragone a good starting  
point for further optimization. 


We designed the initial open Dragone following Granet's method\cite{granet2001}. 
We began with a solution with $F=1.42$, a 1.4~m aperture (that was verified to satisfy the volume constraints), and a large 
DLFOV.  We forced a circular aperture stop 
between the primary and secondary reflectors and numerically optimized its angle and position to obtain 
a larger DLFOV.  The stop diameter was chosen such that for the center feed it projected a 
1.4~m effective aperture onto the primary. 
Adding a stop in this way increased the size of the primary reflector, because 
different field angles illuminated different areas on the reflector.
Actively cooling the aperture stop, however, reduced detector noise, and the stop shielded the 
focal plane from stray radiation. At this stage the system still met 
the Dragone condition and is defined by the `Initial Open-Dragone' parameters in Table~\ref{tab:optics}.

In his publications Dragone provides a prescription to eliminate coma
in addition to the cancellation of astigmatism that is inherent to the baseline designs.\cite{dragone_coma} The corrections 
involve adding distortions to the primary and secondary reflectors 
which are proportional to $r^4$ where $r=0$ is at the chief ray impact point on each reflector. 
We thus attempted to increase the DLFOV using two methods. 

In `Method1', one of the coauthors (RH) used Zemax to add Zernike 
polynomials to the base conics which described the reflectors. These Zernike polynomials were offset from the symmetry axis of the conic 
by $624.2$~cm for the primary and $76.1$~cm for the secondary.  
This placed the origin of polynomials at the chief ray impact point for each reflector. 
Inspired by the Dragone corrections, all Zernike terms up to fourth order and the first fifth order term were allowed to vary. 
The optimization metric was minimization of the rms spot diameter at 
the following locations: the center of the FOV, $\pm2$~deg in Y, and $\pm4$~deg in X. The center of the FOV was given a weight of 100
while each outer point was given a weight of 1. To constrain the optimization, the X and Y 
effective focal lengths were held fixed as was the impact point of the chief ray on the focal plane. 
This optimization step increased the DLFOV
by factors of 1.15, 2.4 and 10.5 at 21, 155 and 799~GHz, respectively. 
We further increased the DLFOV by approximately 50\% at all frequencies by including a curved focal surface and rerunning the optimization.
A small ($\sim$4\%) additional gain in DLFOV was achieved by adding Zernike terms 
up to sixth order, allowing the secondary to focal surface distance to vary, adding a weighted constraint on the effective focal length, 
and adding fields with weight of 0.01 to the rms spot diameter metric at $\pm7.5$~deg in Y and $+15$~deg in X. 
These additional fields were necessary to constrain the corrections at the reflector edges.


\begin{table}[ht]
\centering
\caption{\bf{Telescope geometric parameters  \label{tab:optics}}}
\resizebox{\textwidth}{!}{
\begin{tabular}{|l|llll||ll|}
\hline
\multicolumn{5}{|c||}{PICO optical system}                                    & \multicolumn{2}{c|}{Initial Open-Dragone$^b$}     \\ \hline
                          & Primary           & Secondary    & \multicolumn{2}{c||}{Telescope parameters$^b$} & \multicolumn{2}{c|}{Design parameters}  \\
Reflector size$^a$ (cm)      & $270 \times 205$ & $160 \times 158$ & Aperture (cm)           & 140      & Aperture (cm)                  & 140   \\
Radius of curvature (cm)  & $\infty$         & 136.6             & $F$-number             & 1.42     & $\theta_0$ (deg)           & 90    \\
Conic constant, $k$       & 0                 & -0.926            & h (cm)                    & 624.2    & $\theta_e$ (deg)           & 20    \\
4th ZC$^{c}$ (cm)  & 2018.4            & -61.1           & $\alpha$ (deg)            & 74.2     & $\theta_p$ (deg)           & 140   \\
9th ZC (cm)  & -37.0             & 16.7                & $\beta$  (deg)            &  62.3    & L$_m$ (cm)                     & 240   \\
10th ZC (cm) & -2919.8           & -15.1          & L$_m$ (cm)                &   229.3  &                                &         \\
11th ZC (cm) & -1292.7           & 22.3               & L$_s$ (cm)                &   140.5  & \multicolumn{2}{c|}{Derived parameters} \\ 
12th ZC (cm) & 120.6             & -3.8              &                           &          & $F$-number                     & 1.42  \\   
13th ZC  (cm) & -74.5             & 4.9                &   \multicolumn{2}{c||}{Focal Surface}  & h (cm)                         & 624.2 \\   
19th ZC (cm) & -75.8             & 3.4              & Ellipse major axes (cm)   & 69 x 45  & $\alpha$ (deg)                 & 38.6  \\   
20th ZC (cm) & -398.9            & 6.3            & Ellipse major axes (deg)  & 19 x 13  & $\beta$  (deg)                 & 101.4 \\   
21st ZC (cm) & -319.5            & 23.3              & Radius of curvature (cm)  & 455      & L$_s$ (cm)                     & 122.2 \\   
22nd ZC (cm) & -276.6            & -8.5              &                           &          & Primary, $f$ (cm)              & 312.1 \\   
23rd ZC (cm) & -201.6            & -3.2            &                           &          & Secondary, $a$ (cm)            & 131   \\   
24th ZC (cm) & -127.4            & -1.9            &                           &          & Secondary, $e$                 &  1.802  \\
25th ZC (cm) & -55.0             & 0.1            &                           &          &                                &       \\ \hline
\multicolumn{7}{l}{\footnotesize  $^a$ The maximum physical size of the reflectors.}\\
\multicolumn{7}{l}{\footnotesize  $^b$ Telescope parameters follow the definitions in Granet 2001.\cite{granet2001}} \\
\multicolumn{7}{l}{\footnotesize  $^c$ ZC = Zernike Coefficient; the ZC normalization radius for the Primary (Secondary) mirror is 524.8 (194.1) cm} \\
\end{tabular}
}
\end{table}

In `Method2' another coauthor (JM) used 
CodeV and allowed additional geometric parameters of the system to vary.  To adjust the 
reflector shapes, we added Zernike polynomial corrections to the conic surfaces which defined the two reflectors.
The Zernike polynomials were defined in the same coordinate systems as the base conics.  
We varied the 4th and 9th-13th Zernike coefficients. We allowed the focal surface curvature 
and focal surface to secondary distance L$_s$ to vary.  The primary-secondary distance L$_m$, primary offset $h$, 
and the primary and secondary rotation angles, $\alpha$ and $\beta$, were varied as well.  The optimization 
metric was the rms spot diameter across the field of view, with weighted constraints requiring telecentricity and 
maintaining the X- and Y-focal lengths.  We added Lagrange constraints 
to enforce beam clearances and place an upper limit on overall system size.  Once the optimization converged to 
an acceptable optical system, we added higher order Zernike terms, 19th-25th, and refined 
the reflector shapes using the same metric and constraints. 
The current PICO optical design is from the Method2 optimization procedure.  

\begin{figure} [ht]
\begin{center}
\includegraphics[height=7cm]{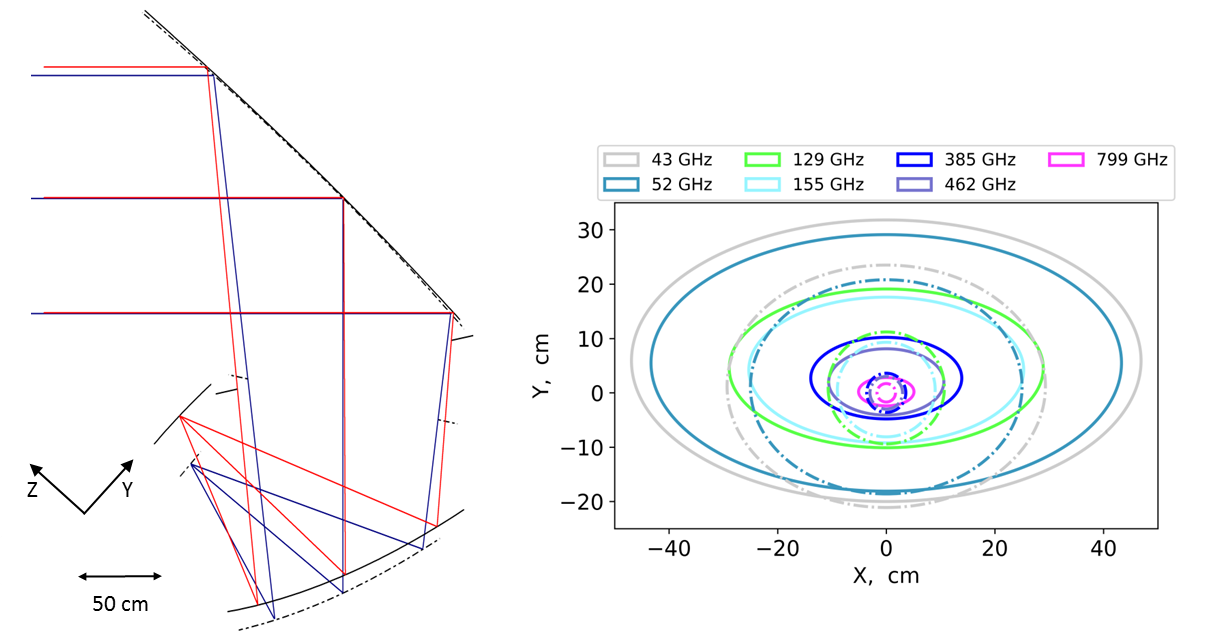}
\end{center}
\caption { \label{fig:compare} 
Comparison between the Method2 open-Dragone and the initial open-Dragone.  
The ray traces (left) are aligned at the chief ray impact point on the primary. 
The final optical system, optimized with Method2 (red rays, solid reflectors), was smaller in the vertical direction 
than the initial open-Dragone (blue rays, dash-dot reflectors). The overlaid Strehl~$=0.8$ contours 
(right) show the improvement achieved with Method2 at all frequencies (solid lines) vs the initial open-Dragone (dash-dot lines) system. 
}
\end{figure}

Figure~\ref{fig:compare} shows that Method2 greatly increased the DLFOV relative to the initial open-Dragone design.  
The DLFOV increased by factors of 1.9, 3.8, 4.3, and 4.6 at 21, 129, 155, and 799~GHz, respectively.  
The most important gain was at frequencies of 129 and 155~GHz. We used this extra area to add 
more C and D pixels, see Figure~\ref{fig:bands} and Section~\ref{sec:focalplane}. The C and D pixels contained the bands 
most sensitive to the CMB, and adding hundreds of these pixels into the focal plane  
gives PICO unprecedented levels of CMB sensitivity. 
Method2 gave somewhat better performance at lower 
frequencies with a DLFOV 1.11 and 1.15 times larger than Method1 at 21 and 155~GHz, respectively.
At 799~GHz Method2 gave DLFOV only 0.3 times that of Method1's area, but the DLFOV still satisfied our science 
requirements. 
Figure~\ref{fig:compare} also shows that Method2 reduced the overall telescope volume and gave more 
physical margin relative to the shadow cone.

The geometric parameters of the PICO optical system are given in Table~\ref{tab:optics}. The 
system is diffraction limited for 799~GHz at the center of the field of view. At 155~GHz the 
DLFOV is 82.4~deg$^2$ and the total throughput at 20~GHz is 910 cm$^2$sr. 
Figure~\ref{fig:strehl} shows Strehl of 0.8 contours for all pixel types.
The slightly concave focal surface, which has a radius of curvature of 4.55~m, is telecentric to 
within 0.12~deg across the entire FOV.

An additional benefit of the optimization is the radius of curvature of the concave focal surface. 
The open-Dragone's focal surface is naturally curved.  Matching this 
curvature reduces defocus, increases the DLFOV, and increases telecentricity.  The unoptimized system was telecentric to within 
2.5~deg while the optimized version is telecentric to within 0.12~deg. If the focal surface was too strongly curved tiling it with flat detector 
wafers would result in large defocus at the edges of these wafers.  This is not the case for PICO. The 4.55~m focal surface radius of curvature, 
results in a defocus of 0.1~mm at the edge of a 10~cm wafer.

\section{FOCAL PLANE}
\label{sec:focalplane}

Modern mm/sub-mm bolometers are photon noise limited. An 
effective way to increase sensitivity is to increase the number of detectors in the focal surface. 
The PICO focal surface has 12,996 detectors, 175 times the number flown on \textit{Planck}. PICO achieves this by 
having a large DLFOV and using multichroic pixels (MCPs)\cite{Suzuki2014_samps,datta2014_mcp}. 
The MCP architecture assumed for PICO has three bands per pixel with two single polarization transition 
edge sensor (TES) bolometers per band and therefore six bolometers per pixel. 
We assume the MCPs were coupled to free space using lenslets and sinuous antennae \cite{Suzuki2014_samps}, 
but the pixel sizes, numbers, and spacing would not change significantly 
if horn or phased array coupling would be used instead.

\begin{figure} [ht]
\begin{center}
\includegraphics[height=6cm]{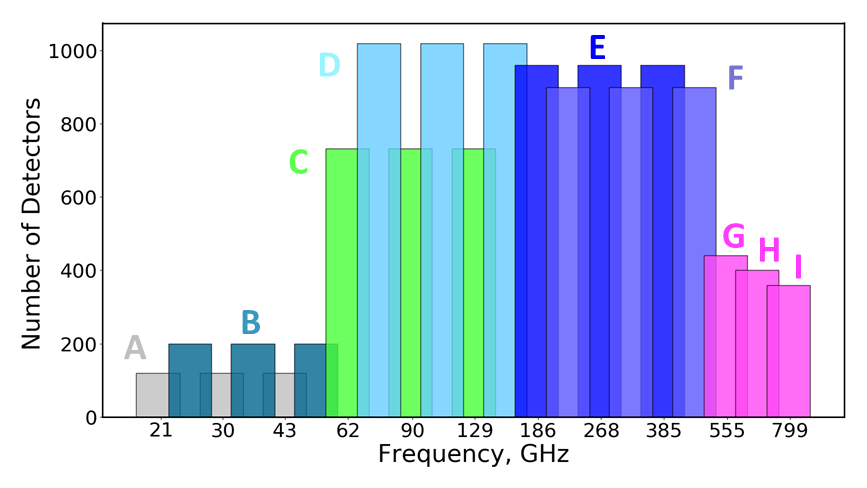}
\end{center}
\caption { \label{fig:bands} 
Frequency coverage of the PICO bands. Each color (excluding magenta) denotes a different MCP, labeled A-F. The bar height 
indicates the number of detectors per band.  Bar width gives the bandwidth. All bands are top-hats with 
25\% fractional bandwidth; the $x$-axis is logarithmic.  The three highest frequencies (magenta) are the 
single color pixels G, H, and I.}
\end{figure}

PICO has 21 overlapping bands with centers spanning the range 21--799~GHz. The bands are divided amongst 
nine pixel types labeled A to I; see Figure~\ref{fig:bands}. 
The 25\% fractional bandwidth is broader than the interband spacing causing neighboring bands to 
overlap and requiring them to be in separate pixels. 
The exceptions to the MCP architecture are the highest three bands, because they are above the superconducting 
band gap of the niobium used for transmission lines, antennae, and filters in MCPs.  
For bands G, H, and I, we assume feedhorn-coupled polarization sensitive bolometers. The technology has high TRL 
as it has been used successfully with 
\textit{Planck}\cite{planck2010_hfi} and Herschel SPIRE\cite{spire2010}.

\begin{figure} [ht]
\begin{center}
\includegraphics[height=7.5cm]{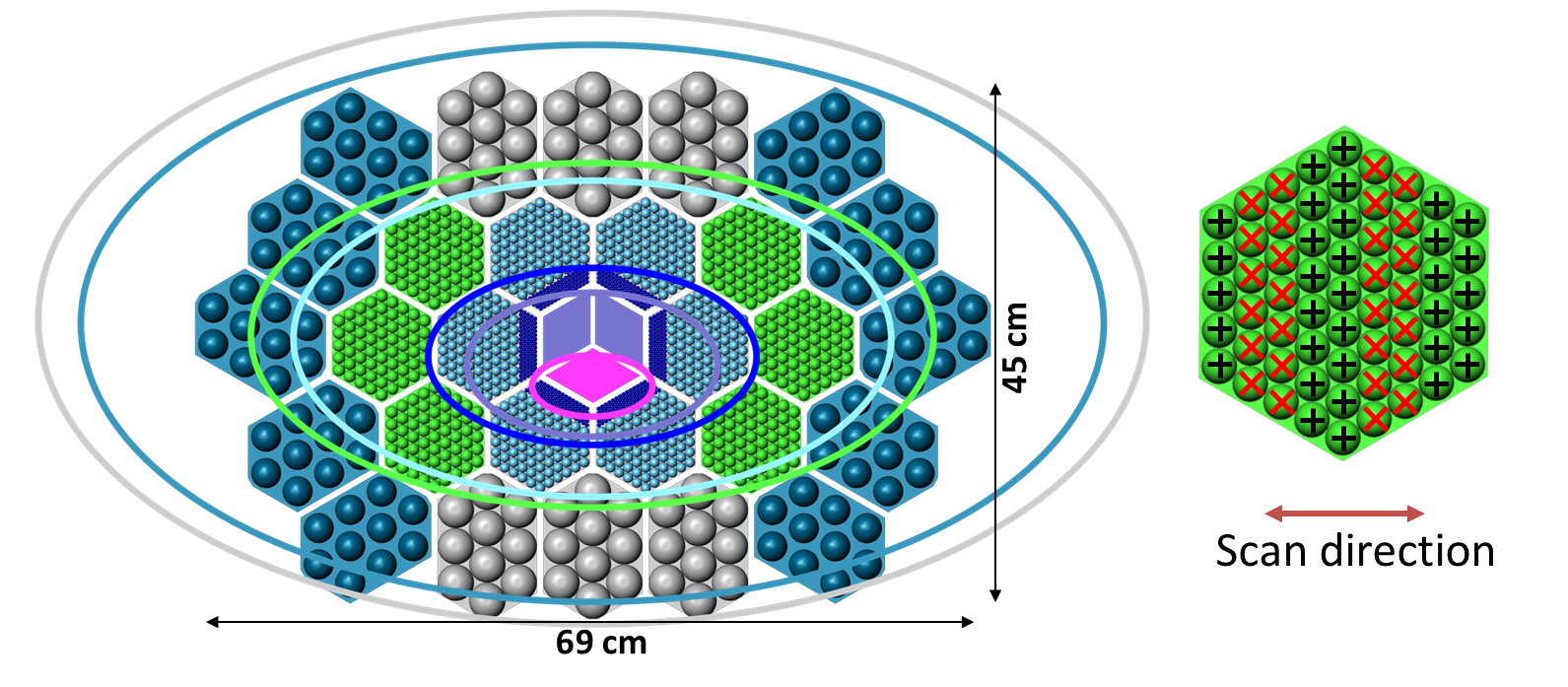}
\end{center}
\caption { \label{fig:focal_plane} \label{fig:QU} 
Left: PICO focal plane layout with Strehl~$=0.8$ contours for each pixel type. The pixel and Strehl contour colors match the band colors, A-I, 
in Figure~\ref{fig:bands}. 
Right: Layout of pixels sensitive to Stokes Q (black crosses) and Stokes U (red exes) for an example wafer.}
\end{figure}

Figure~\ref{fig:focal_plane} shows the PICO focal plane.  
We optimized the diameter of each MCP by calculating the array sensitivity for that pixel type. The calculation included the 
increasing illumination of the aperture stop as the pixel diameter decreased, as described in Section~\ref{sec:det_noise}.
We chose a pixel diameter of $2.1$F$\lambda_{\rm mid}$, where $\lambda_{\rm mid}$ refered to the center 
band of each pixel. This gave an edge taper, $T_e$, on the stop of 10~dB for the center band of each pixel. 

Differencing detectors sensitive to orthogonal polarization states enables each pixel to make a measurement of 
a particular Stokes parameter. Pixels sensitive to the $U$ Stokes parameter are rotated by 45~deg relative to those
that are sensitive to $Q$. This Q/U alignment is in the instrument reference frame with the $x$-axis parallel to the scan 
direction; see Figure~\ref{fig:QU}.


The PICO focal plane readout is designed around $\times$128 time domain multiplexing (TDM), but this choice is not a 
significant driver for the focal plane layout or overall noise budget.

\section{INSTRUMENT NOISE}
\label{sec:noise}

We developed an end to end noise model of the PICO instrument to predict full mission sensitivity and 
provide a metric by which to evaluate mission design trade-offs.  This model assumed white noise
at all frequencies. The overall sensitivity did not include calibration uncertainties or estimates of other possible 
systematic effects. To construct the model we estimated the 
optical load, calculated noise equivalent power (NEP) by source, 
combined all NEP terms to get detector noise, combined all detectors to get noise per frequency band, and then 
included total mission time to find overall mission sensitivity.\cite{suzuki2013_thesis,aubin2013_thesis}  
Each of these steps included various assumptions and design decisions, 
which are discussed in this section.  The assumptions are summarized in Table~\ref{tab:assume}.
The values presented in this section are current best case estimates. Instrument noise {\it requirements}, which would have larger 
noise levels, are still being determined.

To validate our calculations we compared two independent codes that were used for several 
CMB instruments.  The calculations agreed within 1\% for individual noise terms 
and for overall mission noise.  We also used our model to calculate CORE~\cite{core2018_inst} and LiteBIRD~\cite{LB2016_optics,suzuki_private} 
noise from their published system parameters. The 
results were consistent with published values when similar assumptions were used.  

\begin{table}[ht]
\centering
\caption{Noise model parameters, see text for details. }
\label{tab:assume}
\begin{tabular}{|l|l|}
\hline
Throughput                       & single moded, $\lambda^2$          \\
Fractional Bandwidth             & 25\%                                             \\
Reflector emissivity             & $\epsilon = \epsilon_0\sqrt{\nu/\text{150~GHz}}, \epsilon_0 = 0.07\%$ \\
Aperture stop emissivity         & 1                                                \\
Low pass filter reflection loss  & 8\%                                                \\
Low pass filter absorption loss$^a$  & frequency dependent, 0.2\%--2.8\%             \\
Bolometer absorption efficiency  & 70\%                                             \\
T$_e$ of low, middle, and high bands (dB) & 4.8, 10.0, 20.7                                               \\
$\eta_{\rm stop}$ of low, middle, and high bands & 0.68, 0.90, 0.99   \\
Bose noise fraction, $\xi$       & 1                                                \\
Bolometer yield                 & 90\%                                             \\
Bath temperature, $T_o$ (mK)    & 100                                              \\
TES critical temperature, $T_c$ (mK)   & 187                                              \\
Safety factor, P$_{\rm sat}$/P$_{\rm abs}$      & 2                                                \\
Thermal power law index, $n$    & 2                                                \\
Intrinsic SQUID noise (aW/$\sqrt{\text{Hz}}$)   & 3                        \\
TES operating resistance, $\Omega$    &  0.03                               \\
TES transition slope, $\alpha$    & 100                                         \\
TES loop gain                    & 14                                \\
Mission length (years)           & 5                                                \\
Observing efficiency             & 95\%                                             \\
\hline
\multicolumn{2}{l}{\footnotesize $^a$Assumes different thickness per pixel type.}
\end{tabular}
\end{table}

\subsection{Single bolometer noise}
\label{sec:det_noise}

\subsubsection{Model}
\label{sec:model}

\begin{figure} [ht]
\begin{center}
\hspace{1cm} \includegraphics[height=5cm]{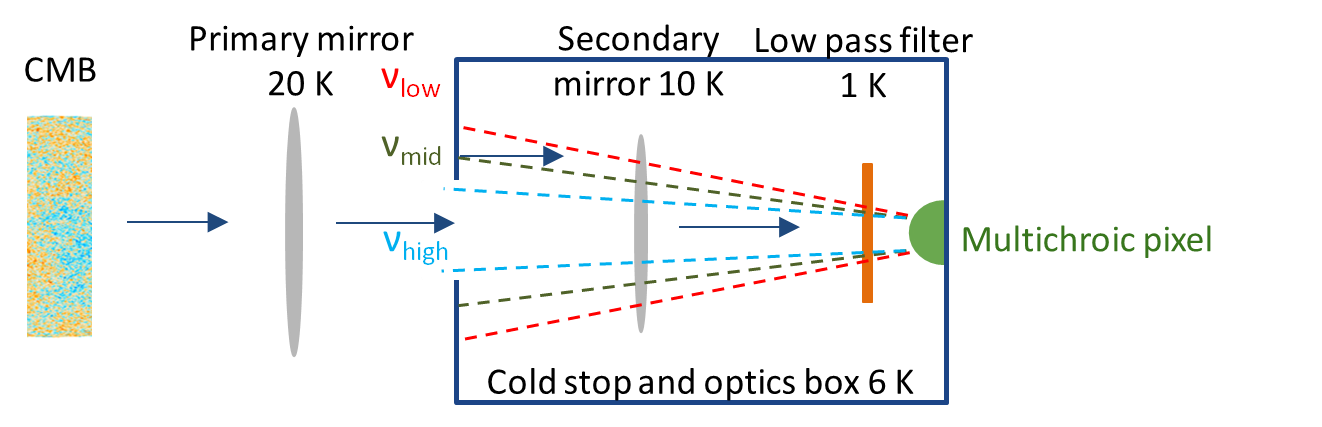}
\end{center}
\caption[load] { \label{fig:load} 
Schematic representation of the prediction of optical load.  Power emitted by each element was modified by the efficiency of 
the following elements and added to the total expected load.  The multichroic pixel illuminated the stop differently for 
each of the three bands. }
\end{figure}

\begin{figure} [ht]
\begin{center}
\begin{tabular}{ccc} 
\hspace{-1.4cm} \includegraphics[height=4.9cm]{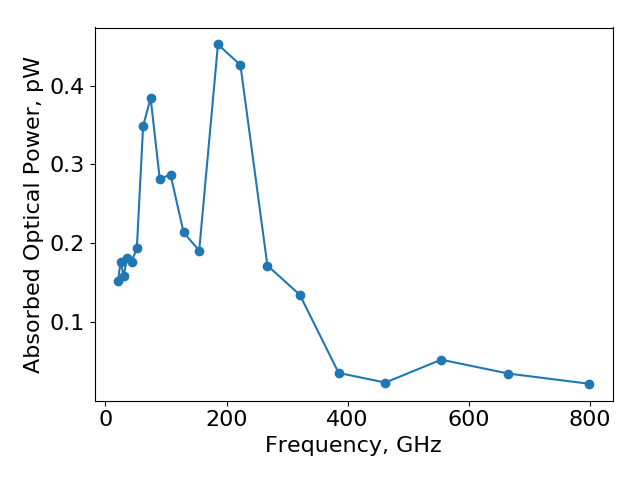} & \hspace{-0.7cm} \includegraphics[height=4.9cm]{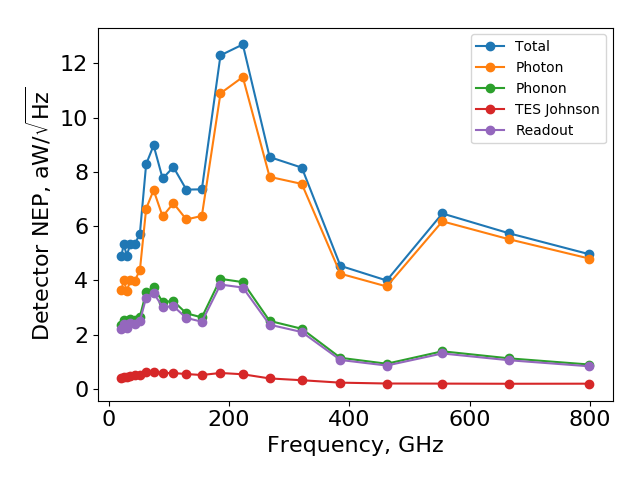} &\hspace{-0.7cm}  \includegraphics[height=4.9cm]{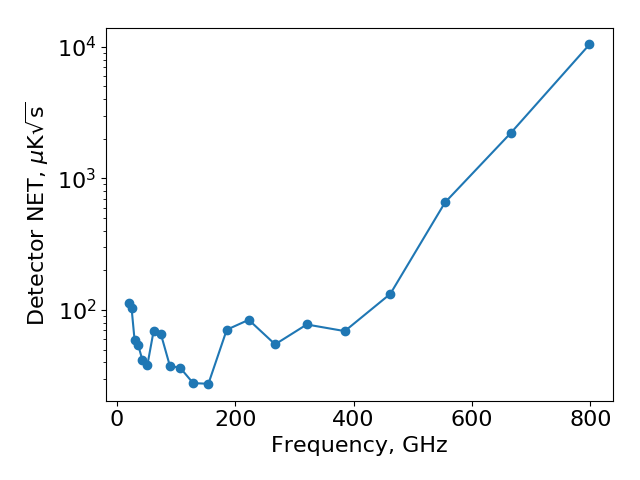} 
\end{tabular}
\end{center}
\caption{ \label{fig:popt} \label{fig:noise} \label{fig:net} 
Left: Expected optical load as a function of frequency for single polarization PICO bolometers. 
Center: Breakdown of NEP sources across the PICO frequency range.  Photon noise dominates even at the lowest frequencies. 
Right: Single detector NET for the PICO bands. 
}
\end{figure}

The sources of optical load are the CMB, the primary and secondary reflectors, the aperture stop, and low pass filters.  
These elements are shown schematically in Figure~\ref{fig:load}. 
The total load absorbed at the bolometer is the sum of the power emitted by each element reduced by the transmission 
efficiency of the elements between the emitting surface and the bolometer.  
The absorbed power is
\begin{equation}
\label{eq:load}
P_{\rm abs} =  (((( P_{\rm CMB} \eta_{\rm PRI} + P_{\rm PRI} ) \eta_{\rm stop} + P_{\rm stop}(1-\eta_{\rm stop}) ) \eta_{\rm SEC} + P_{\rm SEC})\eta_{\rm filter} + P_{\rm filter}) \eta_{\rm bolo},
\end{equation} 
where $P_{\rm elem}$ is the in-band power emitted by a given element for a single polarization and $\eta_{\rm elem}$ is the 
transmission efficiency 
of the element. Power emitted by the stop is a special case. We multiply $P_{\rm stop}$ by 
$(1-\eta_{\rm stop})$ because $\eta_{\rm stop}$ is the spillover efficiency, the fraction of the throughput passing through the 
stop. Therefore $(1-\eta_{\rm stop})$ is the fraction of the throughput which views the stop. We determine 
$\eta_{\rm stop}$ in the following way. 
The MCP angular beam width depends on the wavelength and pixel diameter as\cite{suzuki2013_thesis}
\begin{equation}
\label{eq:mcp_beam}
\theta_{1/e^2} = \frac{2.95 \lambda}{\pi D_{\rm px}}. 
\end{equation} 
We fix $D_{\rm px}$ such that 
the edge taper, T$_e$, of the middle frequency band in each pixel is 10~dB and calculate T$_e$ for the upper and lower bands 
using Equation~\ref{eq:mcp_beam}. This changing illumination of the stop is shown schematically by 
the dashed rays in Figure~\ref{fig:load}. 
For each MCP in pixels A-H, T$_e$ is 4.8, 10, and 20.7~dB for the lower, middle, and upper bands, respectively.  These 
edge tapers correspond to $\eta_{\rm stop}$ of 0.68, 0.90, and 0.99.
The changing $\eta_{\rm stop}$ has two main effects: changing illumination efficiency between bands, which affects optical load 
and therefore the noise equivalent temperature (NET); and telescope beam size not scaling 
smoothly with $\lambda$.
The left panel of Figure~\ref{fig:popt} gives the optical load as a function of frequency. The jumps in load between 
neighboring bands near 70 and 200~GHz, are 
due to $\eta_{\rm stop}$ changing with frequency. 

We consider four noise sources per bolometer; photon, phonon, TES Johnson, and readout. 
Photon noise depends on the absorbed power\cite{richards1994} 
\begin{equation}
\label{eq:photon}
NEP_{\gamma}^2 = \int\limits_{\rm band} 2h\nu p_{\nu} \, d\nu + 2\xi \int\limits_{\rm band} p_{\nu}^2 \,  d\nu,
\end{equation} 
where $p_{\nu}$ is the power spectral density for a single polarization absorbed at the bolometer and $\xi=1$ is the fraction of correlated Bose 
photon noise. We included a factor of 2 in the Bose noise term because the bolometers received power from a single 
polarization. 
Using $P_{\rm abs}$ we calculate the TES bolometer properties and phonon noise.\cite{mather1982}  
We calculate TES Johnson noise, which depends on bolometer bias parameters, for each frequency band.  
All noise sources in the cold and warm readout electronics are lumped under the readout term.  
The combined NEP and the NEP per noise source are shown in 
Figure~\ref{fig:noise} as a function of frequency; the combined values are given in Table~\ref{tab:noise}.

\subsubsection{Results}  

The primary contributor to noise is the optical load; see Figure~\ref{fig:popt}. 
The CMB and stop account for the majority of the optical load 
at all frequencies. Even at 800~GHz the 6.5~K stop
contributes 53\% of the total radiative power compared to 47\% coming from the mirrors.  
The CMB gives more than half the load 
in the middle and upper bands of the multichroic pixels, but the stop dominates the load in the lowest band of each pixel.  
Load from the stop in the lowest band of each pixel ranges from 1.2 times the CMB load at 21~GHz to a maximum of 4.7 
times the CMB load at 223~GHz. 

Bose noise is most significant 
at lower frequencies with NEP$_{\rm Bose}$/NEP$_{\rm Poisson}= 1.5 $ in the lowest band. (All NEP ratios are calculated using 
units of W/$\sqrt{\rm Hz}$ for the dividend and divisor.)  However, Poisson noise increases as 
$\sqrt{P_{\rm abs}\nu}$ while Bose is proportional to $P_{\rm abs}$. Poisson noise equals Bose noise at 30~GHz and 
dominates at higher bands; NEP$_{\rm Bose}$/NEP$_{\rm Poisson} <10\%$ at 321~GHz. 
Phonon noise is the second most significant source, NEP$_{\rm phonon}$/NEP$_{\rm photon}$ ranges from 65\% at 21~GHz 
to 19\% at 799~GHz. 

For TDM readout, phonon and readout noise are approximately equal, and TES Johnson noise is insignificant.  We also modeled 
noise for frequency domain multiplexing readout (FDM).  For FDM the TES Johnson noise is higher, 2/3 of the readout NEP, but the readout 
noise is lower.  Comparing the combined TES Johnson and readout NEPs for TDM and FDM we find essentially identical performance 
with total noise differing by less than 3\% across all bands.  


As mentioned in the beginning of Section~\ref{sec:noise}, the detector NEP we calculate for PICO is consistent with 
that calculated for CORE~\cite{core2018_inst} 
and for LiteBIRD~\cite{LB2016_optics,suzuki_private} when accounting for different input parameters such as the center frequencies, bandwidth,  
and aperture illumination efficiency as a function of frequency. 

\subsection{Combined  array noise}

Using single detector NEPs from Section~\ref{sec:det_noise}, the detector counts from Section~\ref{sec:focalplane}, and 90\% yield for 
each frequency band (see Table~\ref{tab:assume}) we 
calculate the combined NEP of the detector array for each band.  Combining $N$ detectors reduces uncorrelated noise by $\sqrt{N}$.  
This scaling does not hold for Bose photon noise. 
For the lowest band of each MCP the pixel spacing is $0.4$F$\lambda$ and thus the pixels oversample the point spread function 
leading to correlated photon noise between pixels.
Accounting for this effect gives a 26\% increase in the combined array NEP for the 21~GHz band 
and a 0.003\% increase at 799~GHz. 
From the array NEP we convert to noise equivalent temperature (NET) per band using
Numerical values for the NET are given in the second to last column in Table~\ref{tab:noise}.

Assuming evenly weighted observations 
of the full sky, 5 years of mission duration, and 95\% efficiency, we calculate full mission map sensitivities in polarization; see the 
final column in Table~\ref{tab:noise}. 
Combining all bands gives a total CMB map depth for the entire PICO mission of 0.65~$\mu$K$_{\rm CMB}$-arcmin. 

\begin{table}[ht]
\centering
\caption{PICO frequency channels and noise. }
\label{tab:noise}
\begin{tabular}{|c|c|c|c|c|c|c|cc|}
\hline
Pixel  & Band  & FWHM   & Bolometer NEP & Bolometer NET        & N$_{\rm bolo}$ & Array NET         & \multicolumn{2}{|c|}{Polarization map depth}  \\
Type  & GHz   & arcmin    & aW/$\sqrt{Hz}$ & $\mu$K$_{\rm CMB} \sqrt{s}$ &            & $\mu$K$_{\rm CMB}\sqrt{s}$ & $\mu$K$_{\rm CMB}$-arcmin & Jy/sr     \\ \hline
A        & 21  & 38.4 & 4.89   & 112.2   & 120   & 13.6  & 19.1  & 6.69 \\
B        & 25  & 32.0 & 5.33   & 103.0   & 200   & 9.56   & 13.5 & 7.98  \\
A        & 30  & 28.3 & 4.92   & 59.4    & 120   & 5.90   & 8.30 & 7.93   \\
B        & 36  & 23.6 & 5.36   & 54.4    & 200   & 4.18   & 5.88 & 9.59   \\
A        & 43  & 22.2 & 5.33   & 41.7    & 120   & 4.01   & 5.65 & 13.9   \\
B        & 52  & 18.4 & 5.73   & 38.4    & 200   & 2.86   & 4.03 & 16.8   \\
C        & 62  & 12.8 & 8.29   & 69.2    & 732   & 3.14   & 4.41 & 37.0   \\
D        & 75  & 10.7 & 8.98   & 65.4    & 1020  & 2.46   & 3.47 & 48.1   \\
C        & 90  & 9.5  & 7.76    & 37.7    & 732   & 1.49   & 2.09 & 44.5   \\
D        & 108 & 7.9  & 8.18   & 36.2    & 1020  & 1.21   & 1.70 & 57.0   \\
C        & 129 & 7.4  & 7.35   & 27.8    & 732   & 1.08   & 1.53 & 69.7   \\
D        & 155 & 6.2  & 7.36   & 27.5    & 1020  & 0.91   & 1.28 & 84.6   \\
E        & 186 & 4.3  & 12.30 & 70.8    & 960   & 2.51   & 3.54 & 383    \\
F        & 223 & 3.6  & 12.70 & 84.2    & 900   & 3.05   & 4.29 & 579    \\
E        & 268 & 3.2  & 8.55   & 54.8    & 960   & 1.87   & 2.63 & 369    \\
F        & 321 & 2.6  & 8.16   & 77.6    & 900   & 2.73   & 3.84 & 518    \\
E        & 385 & 2.5  & 4.54   & 69.1    & 960   & 2.35   & 3.31 & 318    \\
F        & 462 & 2.1  & 4.00    & 132.6   & 900   & 4.66   & 6.56 & 403    \\
G        & 555 & 1.5  & 6.47   & 657.8   & 440    & 33.1   & 46.5 & 1569  \\
H        & 666 & 1.3  & 5.74   & 2212    & 400     & 117    & 164  & 1960 \\
I          & 799 & 1.1  & 4.97   & 10433   & 360   & 580    & 816  & 2321 \\ 
\hline
Total   &     &      &       &         & 12996 & 0.46   & 0.65 &   \\
\hline
\end{tabular}
\end{table}

\section{CONCLUSIONS/SUMMARY}

The PICO optical system is a two reflector open-Dragone design. It gives a large DLFOV and a total 
throughput larger than 900 cm$^{2}$sr without lenses. The use of a system 
with only reflectors, as opposed to a combination of reflectors and refractors, gives the system high transmission efficiency 
across a wide frequency band and obviates the need to provide broad band anti-reflection coatings. 
Using only two reflectors reduces the volume, mass, and complexity of the system compared to designs with more 
reflectors. The optical system has a cold aperture stop between the primary and secondary reflectors. The stop 
reduces sidelobes while maintaining the other reflectors compact. The native Dragone system has been 
numerically optimized using Zernike polynomials to give a significantly larger DLFOV. 

The focal plane takes advantage of the large DLFOV and MCP 
technology using TES bolometers to implement 12996 polarization sensitive detectors in 21 bands from 21-799~GHz. 
This is the first monolithic CMB instrument to have sensitivity to such a broad frequency range and to baseline a single detector technology 
across this entire band. Under the current set of assumptions, PICO is predicted to have an unprecedented full sky polarization 
map depth of 0.65~$\mu$K$_{\rm CMB}$-arcmin, which is $\sim$80 times the depth {\it Planck} had achieved.



\section{ACKNOWLEDGMENTS}

This Probe mission concept study is funded by NASA grant \#NNX17AK52G.  Gianfranco de Zotti acknowledges financial support from the ASI/University of
Roma--Tor Vergata agreement n.\; 2016-24-H.0 for study activities of the Italian cosmology community. Jacques
Delabrouille acknowledges financial support from PNCG for participating in the PICO study.

\bibliographystyle{spiebib} 
\bibliography{refs} 

\end{document}